\documentclass[conference]{IEEEtran}
\IEEEoverridecommandlockouts

\usepackage{cite}
\usepackage{amsmath,amssymb,amsfonts}
\usepackage{algorithmic}
\usepackage{graphicx}
\usepackage{textcomp}
\usepackage{xcolor}

\usepackage{soul}
\usepackage{subfig}
\usepackage{algorithm2e}
\usepackage{listings}
\usepackage{url}
\usepackage{breakurl}

\usepackage{float}
\usepackage{ctable}

\definecolor{dkgreen}{RGB}{0,64,0}
\definecolor{ltgray}{RGB}{245,245,245}
\definecolor{mauve}{RGB}{139,0,139}

\def\BibTeX{{\rm B\kern-.05em{\sc i\kern-.025em b}\kern-.08em
    T\kern-.1667em\lower.7ex\hbox{E}\kern-.125emX}}

\begin{document}

\title{ML-based Modeling to Predict I/O Performance on Different Storage Sub-systems}

\author{\IEEEauthorblockN{Yiheng Xu\IEEEauthorrefmark{2}, Pranav Sivaraman\IEEEauthorrefmark{2}, Hariharan Devarajan\IEEEauthorrefmark{1}, Kathryn Mohror\IEEEauthorrefmark{1}, Abhinav Bhatele\IEEEauthorrefmark{2}}\\
\IEEEauthorblockA{\IEEEauthorrefmark{2}Department of Computer Science, University of Maryland\\
\IEEEauthorrefmark{1}Center for Applied Scientific Computing, Lawrence Livermore National Laboratory}
}

\maketitle

\begin{abstract}
Parallel applications can spend a significant amount of time performing I/O on large-scale supercomputers. Fast near-compute storage accelerators called burst buffers can reduce the time a processor spends performing I/O and mitigate I/O bottlenecks. However, determining if a given application could be accelerated using burst buffers is not straightforward even for storage experts. The relationship between an application's I/O characteristics (such as I/O volume, processes involved, etc.) and the best storage sub-system for it can be complicated. As a result, adapting parallel applications to use burst buffers efficiently is a trial-and-error process. In this work, we present a Python-based tool called PrismIO that enables programmatic analysis of I/O traces. Using PrismIO, we identify bottlenecks on burst buffers and parallel file systems and explain why certain I/O patterns perform poorly. Further, we use machine learning to model the relationship between I/O characteristics and burst buffer selections. We run IOR (an I/O benchmark) with various I/O characteristics on different storage systems and collect performance data. We use the data as the input for training the model.  Our model can predict if a file of an application should be placed on BBs for unseen IOR scenarios with an accuracy of 94.47\% and for four real applications with an accuracy of 95.86\%.

\end{abstract}

\begin{IEEEkeywords}
I/O performance, trace analysis tool, benchmarking, machine learning, modeling
\end{IEEEkeywords}

\section{Introduction}
Some high performance computing (HPC) applications can suffer from I/O
performance bottlenecks due to slower advances in storage hardware technology
as compared to compute hardware~\cite{gainaru2015schdulingIO}. Such I/O
bottlenecks can have a significant impact on the overall application
performance~\cite{Hjelm2017libhioOI}, thus  people create new I/O subsystems to
improve applications performance. Recently, one type of I/O subsystem that is
gaining more attention is the burst buffer. 

Burst buffers (BBs) are fast intermediate storage layers positioned between
compute nodes and hard disk storage~\cite{teng2016burstbuffer}. Besides doing
I/O all the time on the parallel file system (PFS), applications can let I/O go
through BB and write back to the parallel file system at the end of the job.
Therefore, BBs are suitable for applications with frequent checkpointing and
restart\cite{sato2012checkpointing}. Studies also show that BBs can accelerate
I/O for a variety of HPC applications~\cite{sato2012checkpointing,
nicolae2019veloc, sato2014checkpointing2, bhimji2016nerscBB,
ovsyannikov2016nerscBB}. For instance, Bhimji et al.~\cite{bhimji2016nerscBB}
demonstrate several types of applications including scientific simulation,
data analysis, etc. can achieve better performance on BBs.

\begin{figure}[h]
    \centering
      \includegraphics[width=0.8\columnwidth]{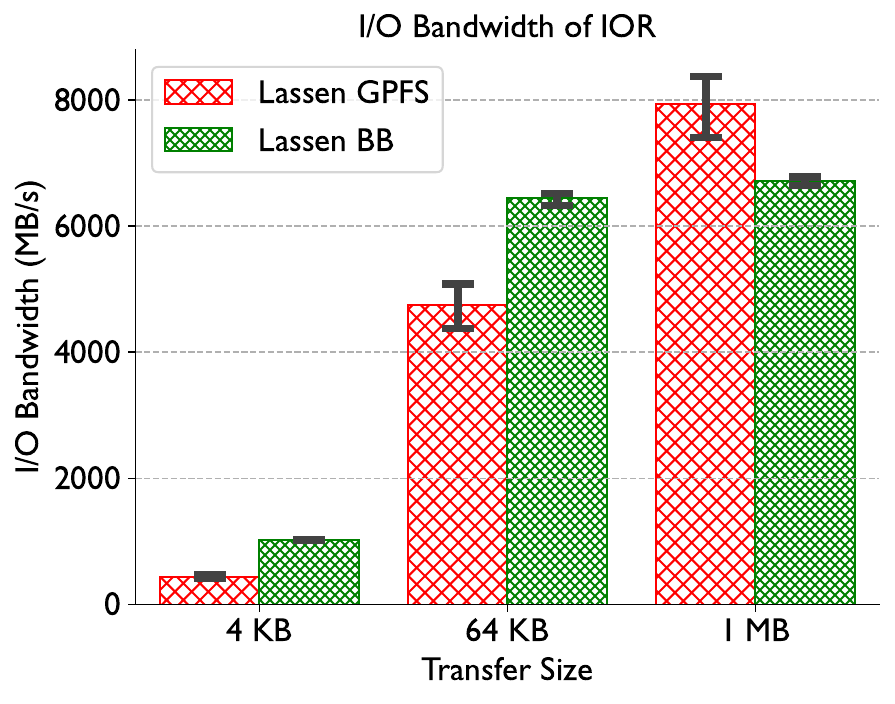}
    \caption{Comparison of I/O bandwidth for different transfer sizes when IOR
uses Lassen GPFS versus burst buffers. Depending on the transfer size, BBs do
not always achieve better I/O performance than GPFS.}
    \label{transfer-size-vs-io-bandwidth}
\end{figure}

Although burst buffers have the potential to improve I/O performance,
determining if an application could be accelerated by putting files on burst
buffers  is a complex task. First, the underlying relationship between an
application’s I/O characteristics and its performance on I/O subsystems can be
considerably complicated. In this paper, we use {\em I/O characteristics} to
refer to things that decide the I/O behavior of an application, such as I/O
volume, I/O library used, etc. For instance, the amount of data transferred per
read or write, which we refer to as {\em transfer size} in the rest of the
paper, can impact the I/O subsystem choices. For a very simple example, we did several IOR~\cite{ior} runs with all configurations the same except different transfer sizes when writing files on
Lassen GPFS (the regular parallel file system on Lassen) and BBs. We find that
with a smaller transfer size it performs better on BB, while with a larger
transfer size it performs better on GPFS (Figure
\ref{transfer-size-vs-io-bandwidth}). Varying other configurations of IOR can
also lead to significant performance differences. We did a suite of IOR runs
with various configurations on both systems, and the most extreme one among them
runs 10x faster on GPFS than on BB. In later sections we present case studies on them and explain such cases through detailed analysis.

Second, even though we can manually figure out whether to direct the I/O of an
application to BBs by doing experiments, it is a trial-and-error process. There
is no work making the I/O subsystem selection an efficient workflow for
arbitrary applications. Moreover, even a single application can read from or
write to multiple files with different I/O characteristics. I/O to some files
can have characteristics that perform better on GPFS, while I/O to other files
can have characteristics that perform better on BBs. Figuring out the placement
of each file manually for large-scale HPC applications is
unrealistic.

Lastly, doing detailed analyses
is challenging as well. Understanding why certain I/O characteristics perform well or poorly
on certain I/O subsystems through detailed analyses is important.
Detailed analyses can inspire insights for optimizations. They also help us
validate our decision of choosing BBs or not. However, doing so is not trivial. Users have to make considerable efforts to write their
own codes for the analysis, as existing tools are not efficient enough
for detailed analysis. They lack of interfaces that enable users to
customize their analysis.

We address the challenges by providing a methodology to conduct detailed
analysis and model the selection of storage sub-systems using machine learning
for HPC applications. We first present our performance analysis tool (PrismIO)
and several case studies we did with it. We identify the bottlenecks of certain
systems and explain why. PrismIO provides a data structure and APIs that enable
users to customize their analysis. It also provides API to extract the I/O
characteristics of an application. Next, we model the relationship between I/O
characteristics and burst buffer selections using machine learning. We run IOR
with various I/O characteristics on several I/O subsystem installations and
collect performance data. With the data, we train models that select the best
I/O subsystem for files of an application based on its I/O characteristics.
Finally, we present a workflow that efficiently extracts I/O characteristics of
applications, executes the model, and gives the file placement plan.

The main contributions of this work are:
\begin{itemize}
\item We build a tool called PrismIO that enables detailed analysis of I/O
traces. We identify and reason about different patterns/bottlenecks of
I/O subsystems.
\item We conduct detailed experiments to study the relationship between I/O
characteristics and the best storage sub-system choices. The resulting dataset
can be used to model this relationship.
\item We train a machine learning model that approximates this relationship on Lassen. The model predicts whether to place
files on BBs or GPFS based on I/O patterns with 94.47\% accuracy.
\item To make it easy to train ML models, we add API functions in PrismIO that
extract I/O characteristics and integrate the modeling workflow into PrismIO.
\end{itemize}

\section{Background \& related work}
In this section, we provide relevant literature needed to understand our work.
We introduce BB, its architecture, and how it benefits I/O performance based on
previous research. We also provide previous works that study how to better use
BB and the limitations they found. Besides, we introduce performance analysis
tools for I/O and their limitations.

\subsection{Burst buffers}

Burst Buffers (BBs) are fast intermediate storage layers between compute nodes
and storage nodes~\cite{teng2016burstbuffer}. The traditional HPC I/O
architecture consists of on-node memory, I/O nodes for handling I/O requests,
and a disk-based parallel file system \cite{guo2008hpcIOarch}. But this
architecture is not enough to fill the gap between computational performance
and I/O performance \cite{bhimji2016nerscBB}. Therefore, BB is a natural
solution to be introduced between these layers. BBs are implemented on several
state-of-the-art HPC clusters such as Lassen, Summit, and Cori.

Burst buffers can be implemented in various ways. In terms of hardware, BBs are
mostly made with SSDs. For example, Lassen at LLNL uses 1.6 TB Samsung PM1725a
NVMe SSDs~\cite{lcsierrasystem}. In terms of architecture, there are two major
types: node-local and shared BBs~\cite{cao2017bbarch}. Lassen and Summit adapt
node-local BBs. In this case, a BB is directly connected to a compute node. All
BBs are independent, which means they don't share the same namespace. For the
shared BBs (as used in Cori), BB nodes are separated from compute nodes and can
be accessed by multiple compute nodes through interconnected
networks~\cite{Landsteiner2016ArchitectureAD}. In this case, compute nodes
perform I/O on the global BB file system with a shared namespace. In this
paper, we focus on node-local BBs.

Research has demonstrated the benefit BB brings to I/O performance
\cite{Hjelm2017libhioOI}. For example, Bhimji et al.~analyze the performance of
a variety of HPC applications on Cori burst buffers, including scientific
simulations, visualization, image reconstruction, and data
analysis~\cite{bhimji2016nerscBB}. They demonstrate that burst buffers can
accelerate the I/O of these applications and outperform regular Parallel File
Systems (PFS). Pottier et al.~\cite{pottier2020modelBB} also demonstrate that
node-local burst buffer significantly decreases the runtime and increases the
bandwidth of an application called SWarp. From the research mentioned above
\cite{Hjelm2017libhioOI, pottier2020modelBB, bhimji2016nerscBB}, it is evident
that BB has the potential to improve the I/O performance of certain HPC
applications.

Although these papers demonstrate BB is promising in improving I/O performance,
most of them raise the same concern: The performance is sensitive to
application I/O patterns~\cite{bhimji2016nerscBB, ovsyannikov2016nerscBB,
pottier2020modelBB}.  Pottier et al.~demonstrated different workflows may not
get the same benefit from BBs~\cite{pottier2020modelBB}. Some of them have
decreased performance compared with the performance on regular file systems.
For instance, they indicated that the shared burst buffer can perform worse
than the regular PFS and is sensitive to the amount of data transferred per
read/write operation.  However, their work is only based on a single
application called SWarp so it might not be general enough to cover common I/O
patterns of HPC application.  Existing research on this issue has not given a
general solution to the problem due to insufficient experiments and modeling to
cover different I/O patterns.  Besides, the metric they use is the total
runtime, which may not be a good I/O metric as it can be significantly affected
by work other than I/O like computation. To summarize, the underlying
relationship between an application’s I/O characteristics and the storage
subsystem choice is still unclear.

\subsection{I/O performance analysis tools}

There exist I/O performance analysis tools that can assist in studying such
relationships. Two popular I/O performance analysis tools are
PyDarshan~\cite{pydarshan, 10.1145/2027066.2027068} and
Recorder-viz~\cite{9150354}. Both of them trace application runs and capture
basic performance information such as time and I/O volume of an operation. They
provide visualization scripts upon their traces. With them users can derive
aggregated I/O metrics such as I/O bandwidth, file sizes, etc.  These tools can
be helpful to start with the I/O subsystems selection research question,
however, they are not efficient enough for detailed analysis due to the lack of
interfaces that enable users to customize their analysis. Users have to make
significant efforts to write their own codes to achieve deeper discoveries.
Besides these two tracing tools, there is an automated analysis project called
Tokio~\cite{tokio1, tokio2}, short for total knowledge of I/O. The project
collects the I/O activities of all applications running on NERSC Cori for a
time period and then provides a clear view of the system behavior. It does a
great job in holistic I/O performance analyses such as detecting I/O contention
and unhealthy storage devices at the system level. However, there is no
application-level analysis so it does not map I/O characteristics to I/O
subsystem selections.

\section{Data collection}
To aid users in the appropriate selection of I/O subsystems to place files in
their application, we conduct detailed analyses. We also leverage machine
learning to model the relationship between I/O characteristics and the best I/O
subsystems. Both of them require good supporting data. Therefore, the
cornerstone of our work is to carefully design an experiment to
build up a large and unbiased dataset. In this subsection, we introduce our experimental
setup and data collection in detail. We go through machines, environments,
benchmark configurations, and how we process the data.

\subsection{HPC machine and storage system}

We conduct our experiment on Lassen. Lassen is a cluster at LLNL using
IBM Power9 CPUs with 23 petaflops peak performance. It uses IBM's Spectrum
Scale parallel file systems, formerly known as GPFS. GPFS is one of the most
common implementations of parallel file systems (PFS). It is used by many of
the world's supercomputers~\cite{GPFS}. In terms of BB, Lassen uses node-local
BBs implemented with Samsung NVMe PCIe SSDs.

\subsection{The IOR benchmark and its configuration}

The benchmark we use is IOR~\cite{ior}. We choose IOR because it is
configurable enough to emulate various HPC workflows. It is proved to be an
effective replacement for full-application I/O benchmarks~\cite{ior-emulation}
\cite{io500}. In our experiment, we select representative I/O characteristics
that are common in real applications and configure IOR with them. For instance,
we include collective MPI-IO because it's a common characteristic in
checkpoint/restart applications. We use IOR to emulate the I/O characteristics
of real applications by covering all possible combinations of the selected I/O
characteristics. We further separate I/O into six types, RAR, RAW, WAR, WAW,
RO, and WO, to better map IOR runs with different types of applications (RAR
for read after read, RAW for read after write, RO for read only, etc.). For
instance, RAR and RO are common in ML applications between epochs. We again
cover all feasible combinations of these read/write types and I/O
characteristics. 

We do the following to take variabilities into account. We repeat each run five
times at different times and days to minimize the impact of temporal system
variability. Since our whole experiment contains around 35,000 runs, we avoid
adding abnormal burdens to the system and reflect system performance as normal
as possible. We run them sequentially so at one time there is only one IOR run.
We collect all I/O metrics from IOR standard output, including total I/O time,
read/write time, bandwidth, etc, and take the mean of five iterations. Each IOR
run produces a sample of the dataset.

\subsection{Data processing}

The data we use is derived from IOR runs with different configurations on
BB and GPFS, where each IOR run is a sample in the dataset. The
independent variables are all I/O features we included in the experiment.
Descriptions of them are presented in Table~\ref{tab:1}. We first process them to make
them suitable for common ML models. Most ML models cannot directly handle
string-formatted categorical columns. Also, if a column is nominal instead of
ordinal, simply converting it into integers ranging from 0 to the number of
categories may result in erroneous inference. One popular solution is using
one-hot vector encoding. It works by creating as many columns as the number of
categories for a categorical column in the data set. Each column represents a
category. We set a column to 1 if the original value is the category
represented by that column and 0 for all other columns.  We encode all our
categorical columns such as file system, API, etc., into one-hot vectors. 

\begin{table}[h]
    \centering
    \caption{Description of data for training the ML models.}
    \label{tab:1}
    \begin{tabular}{ll} \toprule
\textbf{Input I/O feature} & \textbf{Description} \\ \midrule
I/O interface & Categorical feature \\
& Can be POSIX, MPI-IO, HDF5 \\ \hline
collective &  Enable collective MPI-IO \\ \hline
fsync &  Perform fsync after POSIX file close \\ \hline
preallocate & Preallocate the entire file before writing \\ \hline
transfer size & The amount of data of a single read/write \\ \hline
unique dir & Each process does I/O in its own directory \\ \hline
read/write per & The number of read/write \\
open to close & from open to close of a file \\ \hline
use file view & Use an MPI datatype for setting the file \\ 
& view to use individual file pointer \\ \hline
fsync per write & perform fsync upon POSIX write close \\ \hline
I/O type & Categorical feature \\ 
& can be RAR, RAW, WAR, WAW, etc. \\ \hline
random access & whether the file is randomly accessed \\ \bottomrule
\end{tabular}
\end{table}

As our goal is to classify the best I/O subsystem to place a file, the
dependent variable should be the I/O subsystem. Recall that we run every I/O
feature combination on both GPFS and BBs. We obtain the true label by selecting
the I/O subsystem that gives the best I/O bandwidth for each particular I/O
feature combination, which will be either GPFS or BB. The exhaustive I/O feature
combinations plus the true label makes our final dataset. The final dataset has
19 feature columns and 2967 rows.

\section{PrismIO: An I/O performance analysis tool}
We have created PrismIO, a Python library that builds on top of
Pandas~\cite{mckinney:pandas,mckinney:pandas2}, as a solution to better facilitate detailed analysis. It has two primary benefits. First,
it enables users to do analysis on a Pandas dataframe instead of an unfamiliar
trace format. Based on the dataframe structure, PrismIO provides APIs that
assist users in efficiently doing detailed analyses. Second, it provides
functions that extract the I/O characteristics of applications.  It is the
foundation for applying our model to other applications because I/O
characteristics are the input to the model.

\subsection{Data structure}

PrismIO is primarily designed for analyzing Recorder traces. Recorder
\cite{9150354} is an I/O tracing tool. It captures function calls from common
I/O-related libraries such as fwrite, MPI\_file\_write, etc. We choose Recorder over Darshan because trace enables more detailed analysis than profiles. Recorder has a couple of APIs to generate plots, but it's hard for users to efficiently customize analysis. To address this, we implement PrismIO as a Python
library that builds on top of Pandas dataframe to organize data and build APIs
upon it. The primary class that holds data and provides APIs is called IOFrame.
It reorganizes the trace into a Pandas dataframe. Each row is a record of a
function call. Columns store information of function calls, including start
time, end time, the rank that calls the function, etc. It also explicitly lists
information that is non-trivial to retrieve from the complex trace structure
such as read/write access offset of a file.

\subsection{API functions for analyzing I/O performance}

PrismIO provides several analytical APIs that report aggregated information
from different aspects, such as I/O bandwidth, I/O time, etc. It also provides
feature extraction APIs that extract important I/O characteristics users may be
interested in. These APIs will be part of our prediction workflow that we will
discuss in later sections as they extract features needed for the model. In
addition, PrismIO provides visualization of the trace. The following is a
portion of frequently used APIs. Due to the space limit, we do not discuss all
APIs in detail. Readers can check our GitHub repo for details if interested.

\begin{figure}[h]
    \centering
        \includegraphics[width=0.5\columnwidth]{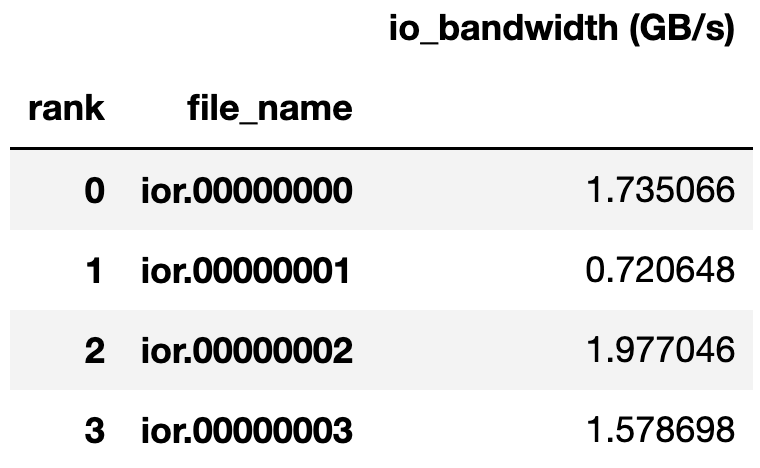}
\begin{lstlisting}
ioframe = IOFrame.from_recorder("/path/to/recorder/trace")
ioframe.io_bandwidth(rank=[0,1,2,3], filter=lambda x: x["I/O_type"] == "read")
\end{lstlisting}
    \caption{I/O bandwidth of ranks 0--3 for each file they read (in a sample
IOR trace.) It returns a dataframe with a hierarchical index that shows the
read bandwidth of rank 0-3 to different files. It utilizes the filter option to
only report read bandwidth. The function automatically selects the most
readable unit for bandwidth, in this case GB/s.}
    \label{io_bandwidth}
\end{figure}

\vspace{0.07in}
\noindent{\textbf{io\_bandwidth}}
One common analysis for I/O performance is to check the I/O bandwidth of each
rank on different files. This function groups the dataframe by file name and
rank and then calculates the bandwidth by dividing summed I/O volume by summed
I/O time in each group. It provides options for quick filtering. It is common
that people only want to focus on certain ranks in parallel application
analysis. We provide a ``\texttt{rank}" option that takes a list of ranks and
filters out ranks that are listed before the groupby and aggregation. We also
provide a general filter option where users can specify a filter function. The
following code and Figure~\ref{io_bandwidth} demonstrate an example of using
\texttt{io\_bandwidth}.

\begin{figure}[h]
    \centering
        \includegraphics[width=0.95\columnwidth]{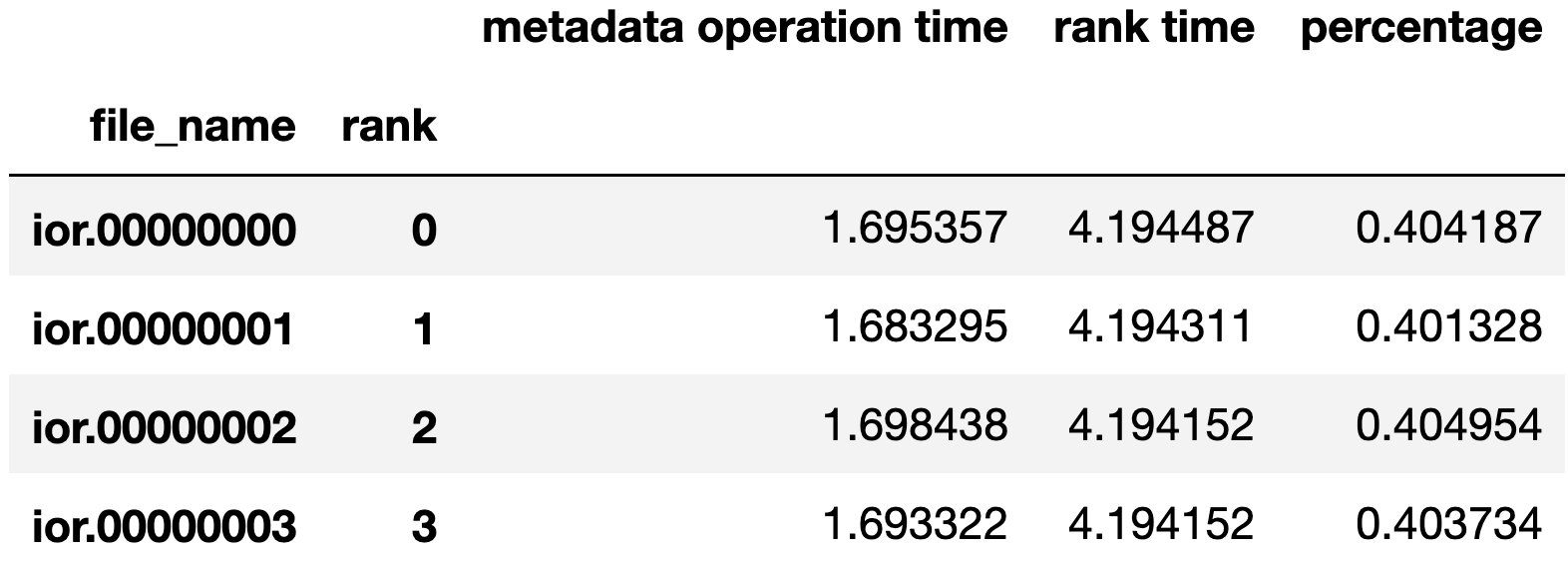}
\begin{lstlisting}
ioframe.io_time(rank=[0,1,2,3], filter=lambda x: x["I/O_type"] == "meta")
\end{lstlisting}
    \caption{Time spent in metadata operations by ranks 0--3 for each file they
read (in a sample IOR trace.) In addition to absolute time, it also reports the
percentage of total time spent by a process performing I/O.}
    \label{io_time}
\end{figure}

\vspace{0.07in}
\noindent{\textbf{io\_time}}
\texttt{io\_time} returns the time spent in I/O by each rank on different
files. Similar to \texttt{io\_bandwidth}, users can filter different things.
Figure~\ref{io_time} demonstrates an example of using it to check how much time
rank 0, 1, 2, 3 spent in metadata operations (open, close, sync, etc.) on each
file.

\begin{figure}[h]
    \centering
        \includegraphics[width=0.95\columnwidth]{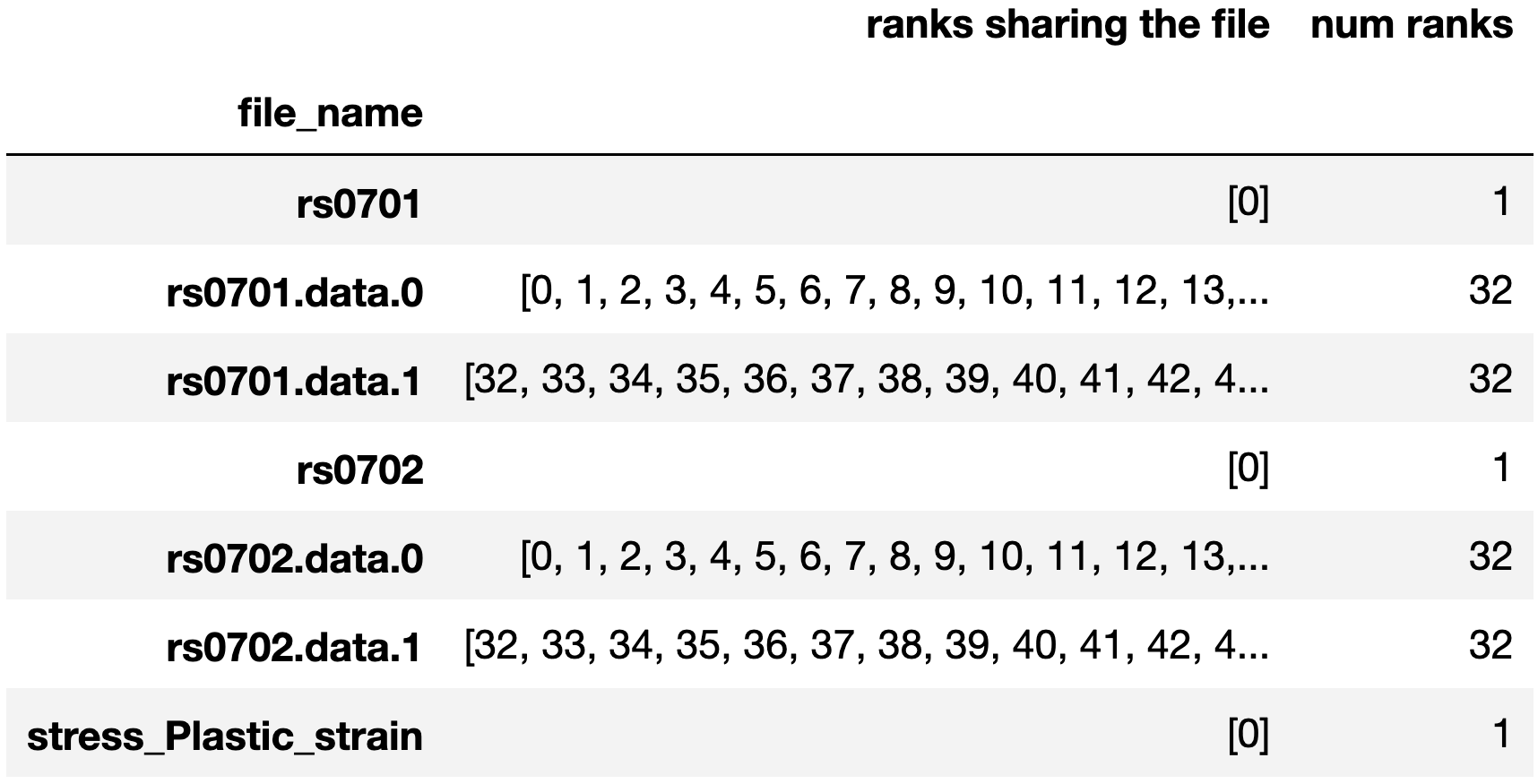}
\begin{lstlisting}
ioframe.shared_files(dropna=True)
\end{lstlisting}
    \caption{A screenshot of a part of the DataFrame when using shared\_files for a sample trace. It demonstrates how files are shared across processes. Users can easily observe some files are not shared
    and some files are shared by 32 ranks.}
    \label{shared_files}
\end{figure}

\vspace{0.07in}
\noindent{\textbf{shared\_files}}
Another common analysis about parallel I/O is to see how files are shared
across ranks.  \texttt{shared\_files} reports for each file, how many ranks are
sharing it and what those ranks are. Figure \ref{shared_files} demonstrates
part of the result of \texttt{shared\_files} of a run. 

\subsection{Feature extraction API}

Feature extraction is a class of APIs that extract the I/O characteristics of
an application. As introduced previously, one primary purpose of these APIs is
to prepare inputs to the model. In the machine learning context, the inputs to
a model are called features. As our model models whether to place files on BBs
based on I/O characteristics of applications, I/O characteristics are the input
features to our model. In the rest of the paper, when we discuss topics related
to the model, we use I/O features to refer to I/O characteristics. Such
features include I/O type (RAR, RAW, WAR, etc.), access pattern (random or
sequential), file per process (FPP), I/O interface (POSIX, MPI-IO, etc.), etc.
For an arbitrary application, users either need to read the source code or
analyze the trace to get them. With the original trace, users may spend
considerable time writing code to analyze it, while with PrismIO, users can get
those features by function calls. We provide examples to demonstrate the
simplicity of using them.

\begin{figure}[h]
    \centering
        \includegraphics[width=0.8\columnwidth]{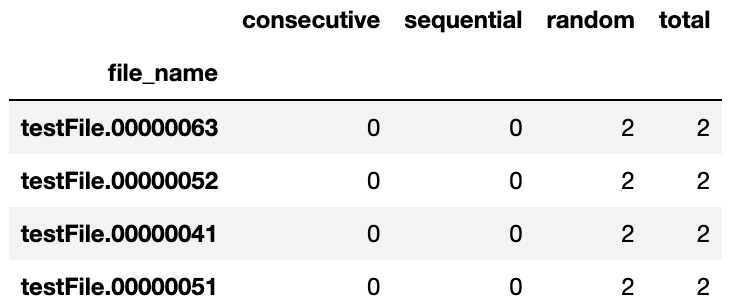}
\begin{lstlisting}
ioframe.access_pattern()
\end{lstlisting}
    \caption{A screenshot of a part of the DataFrame when using access\_pattern
for a sample trace. It counts the number of different access types. It has 2
random accesses on files and no other type of access. From the count users can
know what kind of access the run is mostly doing to files, and thus decide the
access pattern for files.}
    \label{access_pattern}
\end{figure}

\vspace{0.07in}
\noindent{\textbf{access\_pattern}}
\texttt{access\_pattern} counts the number of different access patterns
including consecutive, sequential, and random access. Figure
\ref{access_pattern} demonstrates the \texttt{access\_pattern} output of an
example trace.

\begin{figure}[h]
    \centering
        \includegraphics[width=0.8\columnwidth]{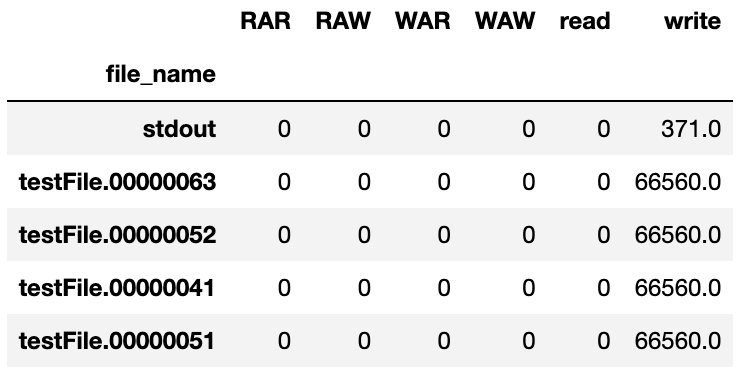}
\begin{lstlisting}
ioframe.read_write_pattern()
\end{lstlisting}
    \caption{A screenshot of a part of the DataFrame when using
readwrite\_pattern for a sample trace. It calculates the amount of I/O for
different read/write patterns. This example run only does write so the pattern
of all files is write only (WO).}
    \label{readwrite_pattern}
\end{figure}

\vspace{0.07in}
\noindent{\textbf{readwrite\_pattern}}
\texttt{readwrite\_pattern} identifies read/write patterns such as RAR, RAW,
etc., and calculates the amount of data transferred with those patterns. Figure
\ref{readwrite_pattern} demonstrates the output of \texttt{readwrite\_pattern}
of a sample trace.

\subsection{Visualization API}

\begin{figure}[h]
    \centering
        \includegraphics[width=0.95\columnwidth]{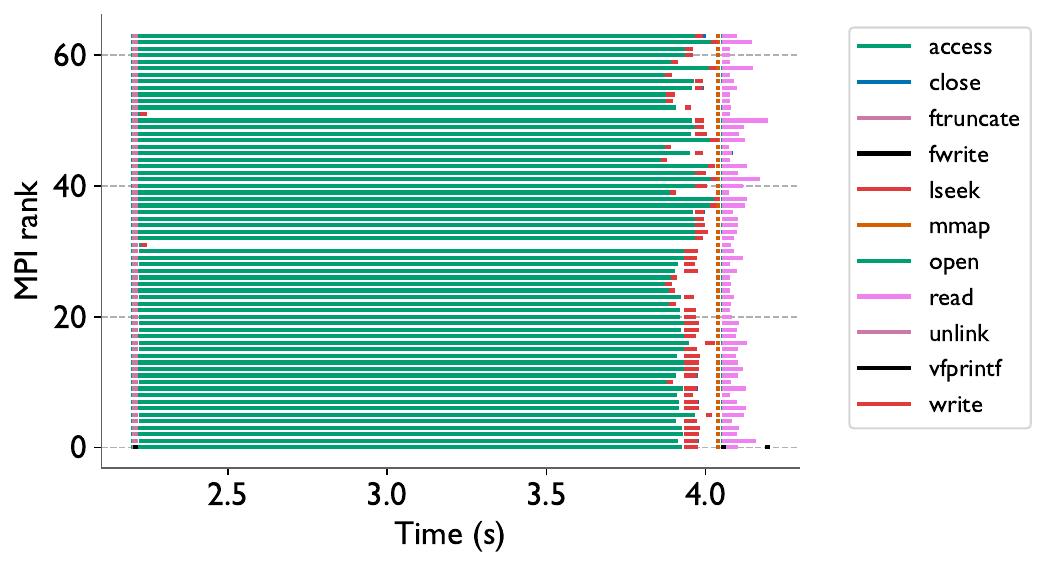}
\begin{lstlisting}
ioframe.timeline(filter=lambda x: x["function_type"] == "I/O")
\end{lstlisting}
    \caption{Timeline of I/O functions of a sample trace. The x-axis is the
time, and the y-axis is the rank ID. Users can easily make observations such as
functions that are abnormally slow, load imbalance, temporal orders of function
calls, etc. For this example, users can easily identify that \texttt{access}
runs slow on most ranks, but fast on two specific ranks. Users can efficiently
get important insights and get directions for further analysis.}
    \label{timeline}
\end{figure}

\vspace{0.07in}
\noindent{\textbf{timeline}}
The most-used visualization API in our use case study is \texttt{timeline}. It
plots the timeline of function calls on each rank. Each line covers the time
interval of a function execution, from start to end. It provides filter options
for users to select certain types of functions. Figure \ref{timeline}
demonstrates an example timeline plot of I/O functions for a sample trace. 

\section{Case studies}
In this section, we present interesting case studies to demonstrate our
analysis with PrismIO. We present insights that have not been systematically
studied via large-scale experiments.

\begin{figure*}[t]
  \centering
\subfloat[Lassen GPFS on 2 nodes]{%
      \includegraphics[width=0.33\linewidth]{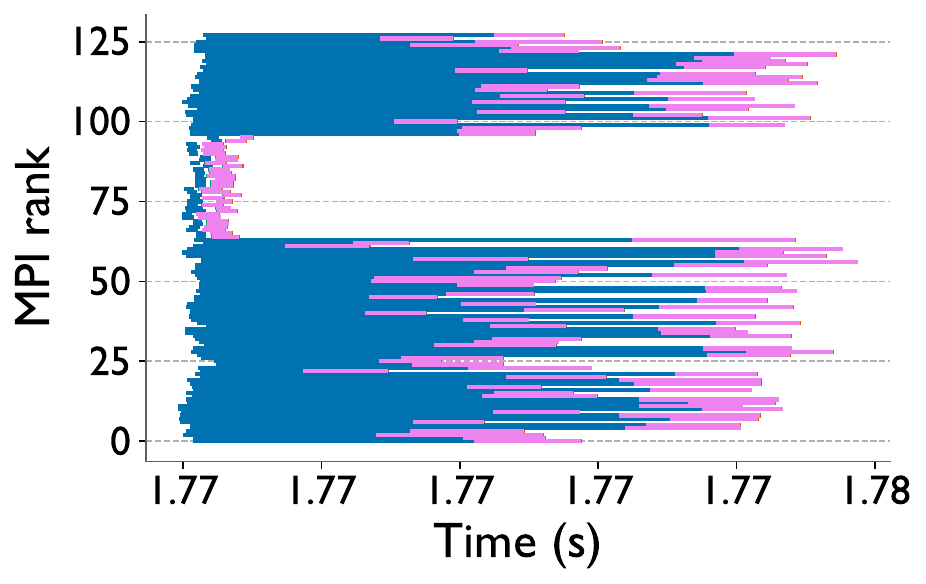}}
  \hfill
\subfloat[Lassen GPFS on 4 nodes]{%
      \includegraphics[width=0.33\linewidth]{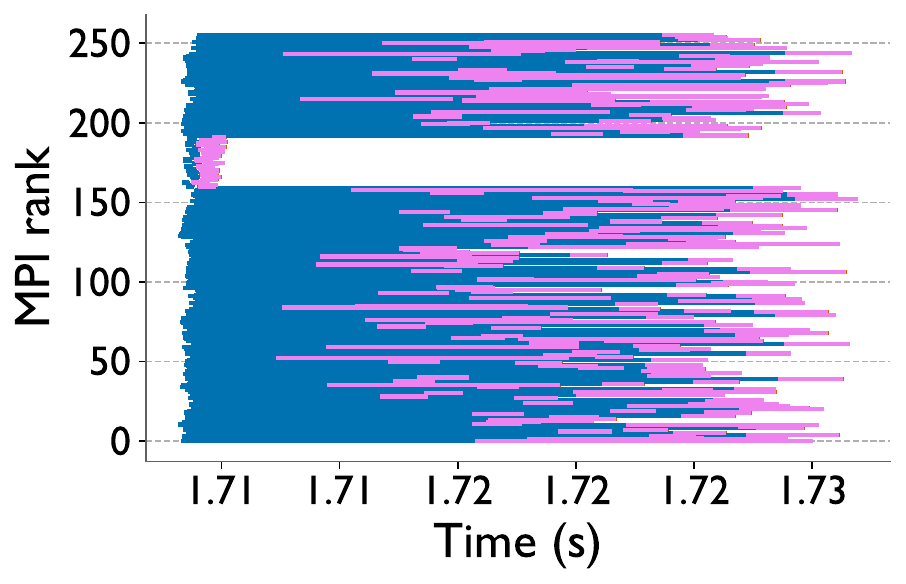}}
   \hfill
 \subfloat[Lassen GPFS on 8 nodes]{%
       \includegraphics[width=0.33\linewidth]{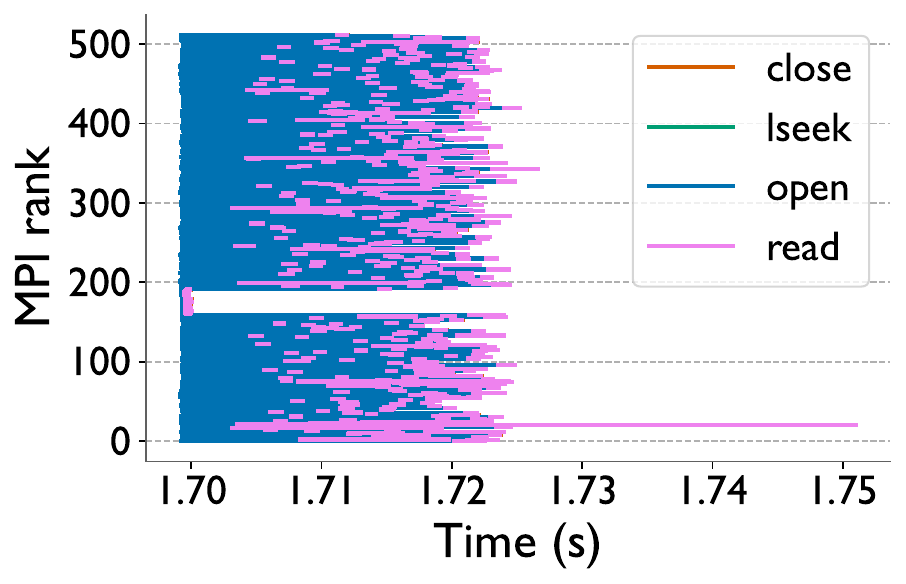}}
\caption{I/O timeline of IOR runs with ``Good BB Bad GPFS" configuration on
Lassen on 2, 4, 8, and 16 nodes on GPFS. Each node holds 32 processes. No
matter how many nodes we use, there is one and only one node that performs much
better than other nodes.}
\label{gpfs-imbalance}
\end{figure*}

\subsection{Extreme runs analysis}

\begin{figure}[h]
  \centering
     \includegraphics[width=0.95\linewidth]{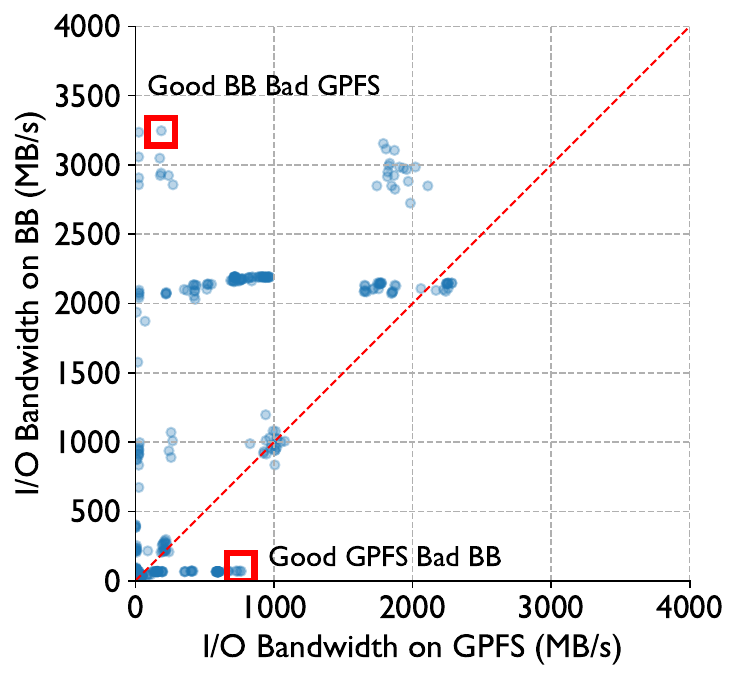}
\caption{I/O bandwidth of IOR benchmarking runs on PFS vs BB. Each point
represents a run with a certain I/O feature combination. The x-axis is the I/O
bandwidth on PFS and the y-axis is the I/O bandwidth on BB. The value is the
average over 5 iterations of that I/O configuration as mentioned in section IV.
The red boxes highlight the most extreme runs we picked for deeper analysis.}
\label{extreme-point}
\end{figure}

We use IOR runs that perform quite differently between PFS and BB from the
experiment dataset introduced in Section IV as our use cases for analysis and
call them extreme runs. Figure \ref{extreme-point} plots the I/O bandwidth of
IOR runs with 64 processes on 2 nodes on GPFS and BB on Lassen. Each
point represents a run with a certain I/O feature combination. The x-axis is
the I/O bandwidth on PFS and the y-axis is the I/O bandwidth on BB. The farther the point from the diagonal, the IOR run with that
configuration performs more differently on PFS and BB. 

We pick the highlighted runs in Figure \ref{extreme-point} to conduct detailed
analyses. The upper left one performs much better on BBs than on PFS. The lower
middle one performs much better on PFS than on BB. We refer to them as ``Good
BB Bad PFS" and ``Good PFS Bad BB" in later text. Since they perform very
differently between systems, they may indicate bottlenecks of systems. We trace
them with Recorder, and then do detailed analyses with PrismIO. 

\begin{figure}[h]
  \centering
\subfloat[Lassen GPFS]{%
     \includegraphics[width=0.5\linewidth]{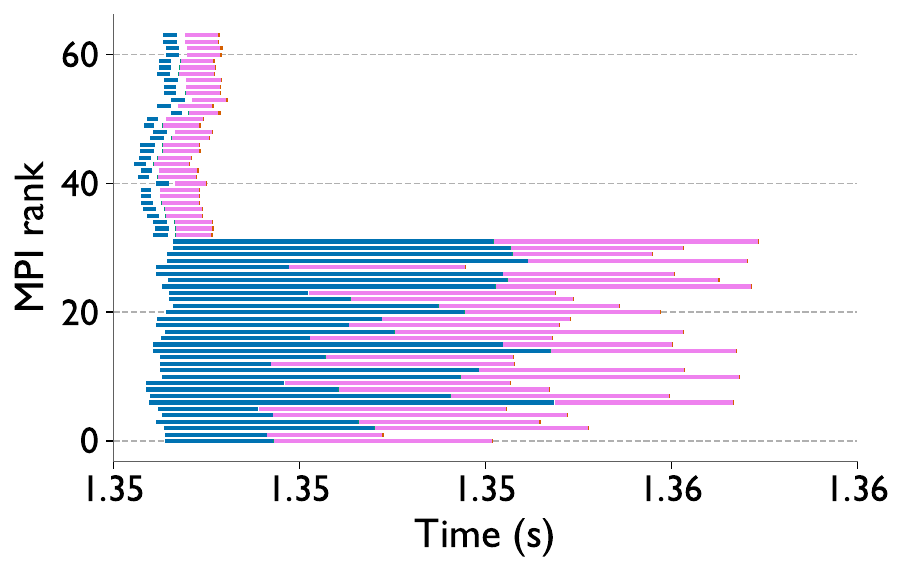}}
  \hfill
\subfloat[Lassen BB]{%
      \includegraphics[width=0.5\linewidth]{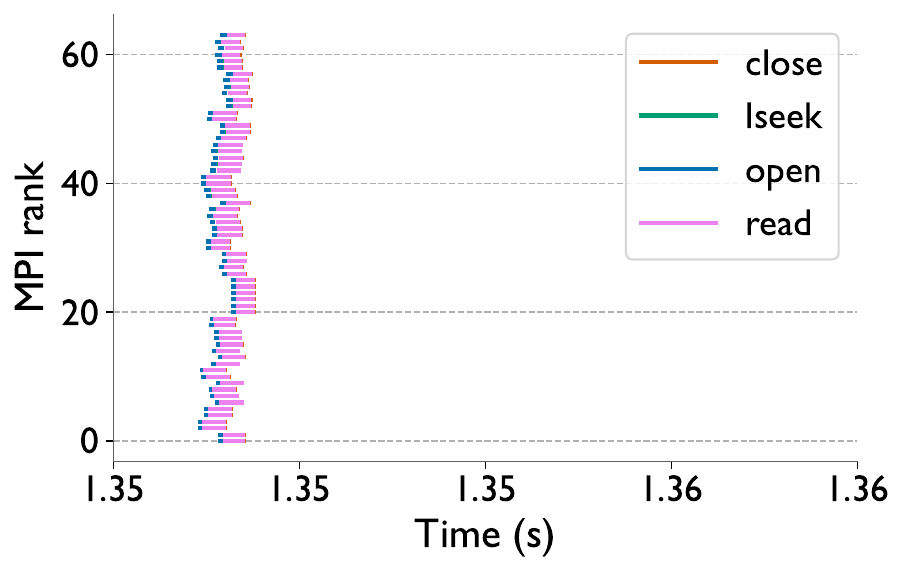}}
\caption{Timeline of I/O function calls of IOR runs on GPFS and BB using the
configuration from the extreme point ``Good BB Bad GPFS". We run them on two
nodes and 32 processes per node. The X-axis is the timeline, and the y-axis is
the rank ID. Each line covers the running time interval of a function from
start to end.}
\label{lassen-good-bb-bad-gpfs}
\end{figure}

\vspace{0.08in}
\noindent{\bf Lassen GPFS bottleneck analysis:}
We first present the analysis for the “Good BB Bad PFS” run on Lassen. The I/O
bandwidth per process for that run on GPFS is 155.01 MB/s, whereas on BB it is
3236.24 MB/s. Such a significant gap indicates there must be performance issues
on GPFS for this run. We first utilize PrismIO \texttt{io\_time} API and find
the run on GPFS spends 37.8\% of the I/O time in metadata operations. We then
use the PrismIO \texttt{timeline} API introduced in section III to visualize
the I/O function traces (Figure \ref{lassen-good-bb-bad-gpfs}). We immediately
observe a significant load imbalance of open over nodes on GPFS: the 32
processes on one node run much slower than 32 processes on the other node, but
this is not happening on BB. We have checked the source code of IOR and
logically all processes do exactly the same thing, so it is likely an issue
with the system. To further drill down, we check if it is because that
particular node is problematic or if it is a system issue.

\begin{figure*}[t] 
  \centering
\subfloat[Lassen GPFS]{%
     \includegraphics[width=0.33\linewidth]{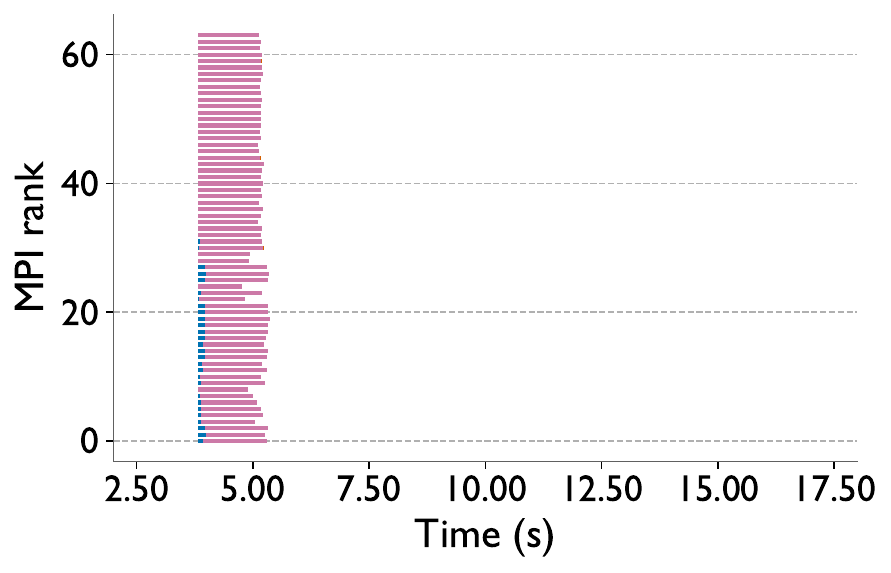}}
  \hfill
\subfloat[Lassen BB]{%
      \includegraphics[width=0.33\linewidth]{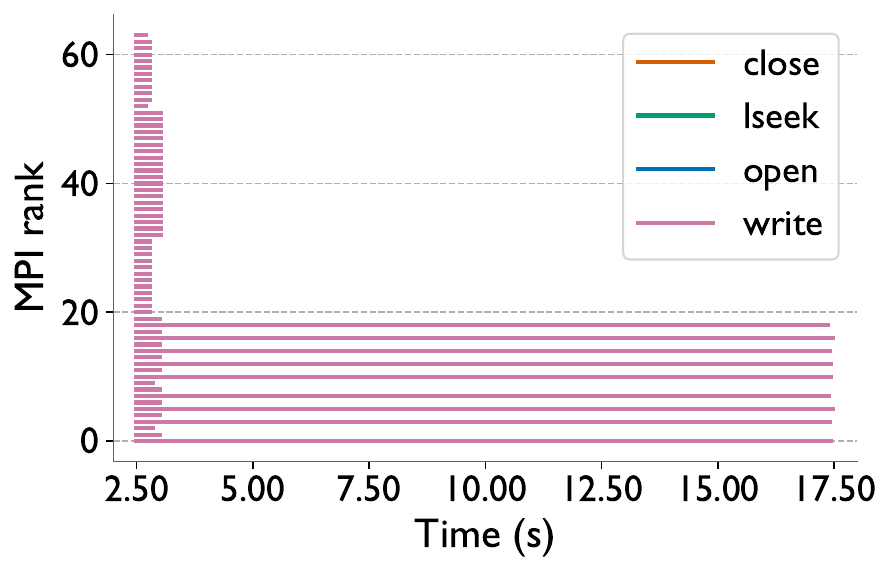}}
  \hfill
\subfloat[Lassen BB with direct I/O]{%
      \includegraphics[width=0.33\linewidth]{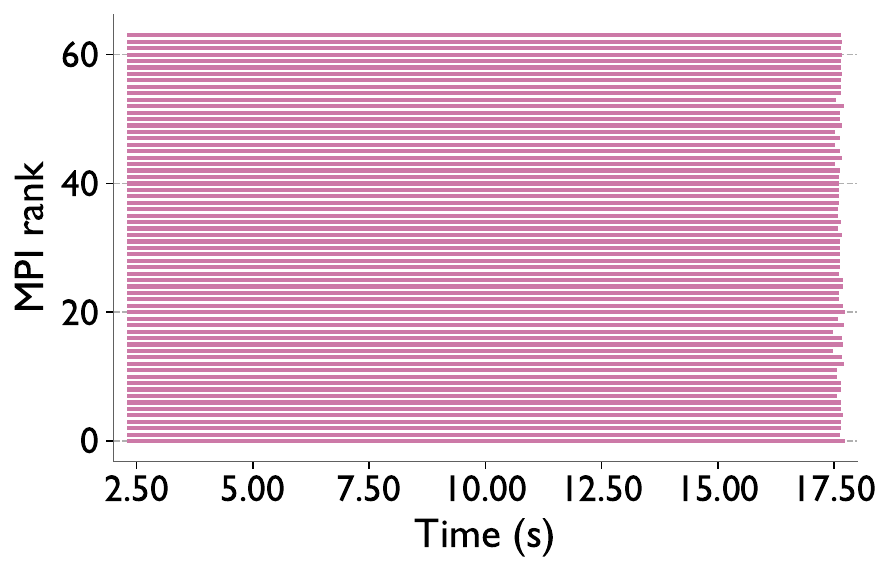}}
\caption{Timeline of I/O function calls of IOR runs on Lassen GPFS and BB using
the configuration from the extreme point ``Good GPFS Bad BB". We run them on
two nodes and 32 processes per node. }
\label{lassen-good-gpfs-bad-bb}
\end{figure*}

We run IOR with the same configuration but on more nodes. Figure
\ref{gpfs-imbalance} demonstrates the I/O timeline of IOR runs on 2, 4, 8, 16
nodes. We noticed there is one and only one node that performs much better than
the other nodes, so we checked if it is the same one in these four runs.
Unfortunately, we did not observe a common node.  Therefore, we conclude that
it's a system problem of GPFS in balancing metadata operations like open. We
checked with LC admins and confirmed the low-level reason is that GPFS does not
have metadata or data servers. When writing a bunch of files to the same
directory, one of the nodes becomes the metadata server for that directory and
thus ranks on it are much faster than ranks on other nodes for metadata
operations.

\vspace{0.08in}
\noindent{\bf Lassen BB bottleneck analysis:}
Similar to the previous section, we dive deeper to analyze the bottlenecks of
Lassen BBs based on the “Good GPFS Bad BB” run. The run is running on 2 nodes
with 32 processes per node as well. Figure \ref{lassen-good-gpfs-bad-bb} (a)
(b) demonstrates the timeline for I/O function calls of each rank in that run
on GPFS and BB, respectively. This time we observe a significant load imbalance
over ranks on BB. Some ranks perform much worse than others, and such ranks are
spread across both nodes. Besides, faster ranks on BB actually perform better
than on GPFS. Furthermore, we noticed those faster ranks have unexpectedly high
bandwidths, which indicates they are probably taking advantage of the SSD
hardware cache. As it is part of the BB architecture, in most real-world use
cases, applications have to use it by default. It does not make sense to
exclude it from I/O performance modeling and analysis. To 
confirm this hypothesis, we run these IOR runs with the same configurations but
with direct I/O enabled, in which case all I/O traffic would skip any kind of cache and
directly go to storage.

Figure \ref{lassen-good-gpfs-bad-bb} (c) demonstrates the I/O timeline of the
same run on BB but with direct I/O enabled.  All ranks now take the same long
time on BBs. This confirms that the load imbalance over ranks on BBs is due to
the SSD hardware cache. GPFS is a shared file system so its hardware cache is
likely to be occupied by millions of other workloads, while the hardware cache
of node-local BBs is not filled until the total amount of I/O exceeds the cache
size. When the hardware cache is filled, ranks with I/O that cannot use the
cache have to run much slower than ranks with I/O that can use the cache.
Therefore, the empirical conclusion from some research that large I/O does not
perform well on BB is not completely correct. A single large I/O can still
perform better on BB than on GPFS if the BB hardware cache is not full. Users
can still do large I/O on BB, as long as the total amount of I/O at a time does
not overwhelm the hardware cache.

\subsection{Overall trends}

\begin{figure}[h]
  \centering
  \subfloat[transfer size]{%
    \includegraphics[width=0.5\linewidth]{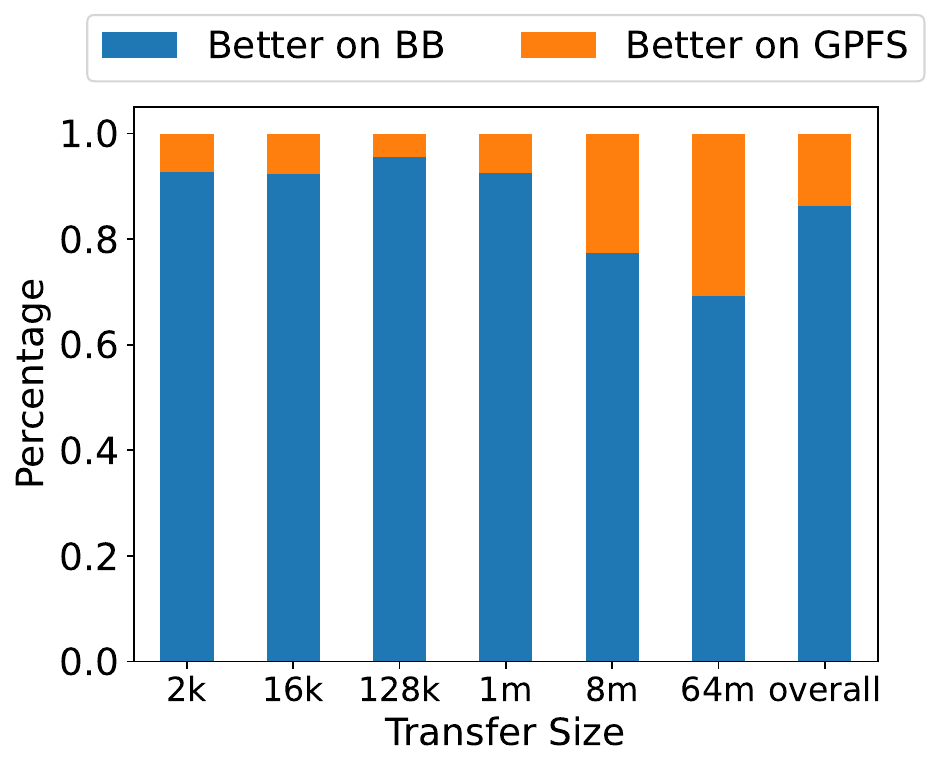}}
   \hfill
  \subfloat[I/O pattern]{%
    \includegraphics[width=0.5\linewidth]{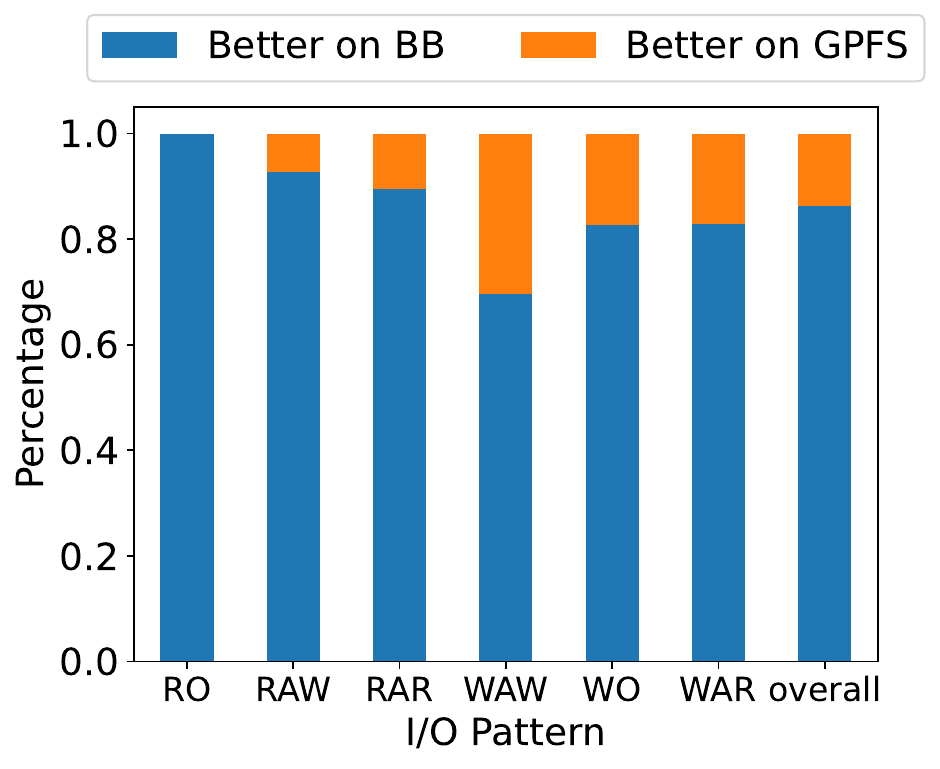}}
\caption{Percentage of runs better on Lassen BB 
given different transfer sizes and I/O operations. }
\label{better-bb-percentage-transferSize}
\end{figure}

From the previous case studies, we identified some values of features that can
be more preferable to BB. We saw MPI-IO, POSIX, transfer size, etc. can
affect the choice of BB. Moreover, later in the model section, we get feature
importance from the model. We will observe transfer size and read/write
patterns are important features for the prediction. To give a summary of such
trends, for a certain feature, we fix its value, and look at all runs with that
feature equal to that value. We count how many runs are better on BB and PFS.
Figure \ref{better-bb-percentage-transferSize} demonstrates the percentage of
runs better on Lassen BB as the values of features change. As transfer size grows, the percentage of runs better on BB decreases. Similarly, 
the percentage of runs better on BB is higher when reading than writing. We can conclude that
although overall BB brings better performance than GPFS, it prefers
smaller transfer sizes and read operations. 

\section{ML model for selecting best I/O sub-systems}
We apply machine learning to efficiently recommend which file should be placed
on which system, which we refer to as ``file placement" in the rest of the
paper.  We design and train an ML model that selects the best I/O subsystems
for application file placements. Moreover, to efficiently apply the model to an
application when users don't know its I/O characteristics, we make the
selection process into a workflow with assistance from PrismIO feature
extraction. In this section, we discuss the design of our models and the
workflow.

\subsection{Model design}

The key to answering whether to place files written by an application on burst
buffers is to understand the relationship between the I/O characteristics of an
application and it's performance on different I/O subsystems. As defined previously, we use I/O features to refer to I/O
characteristics that we use as inputs to our model. Since some parameters are continuous such as transfer size and number of read/write, we need a model to interpolate the relationship. And since the parameter space
is large, heuristic-based methods may not work well and are difficult to be extended
when adding new I/O features. Therefore, we decided to apply machine learning
(ML) techniques to model such relationships. An application can have multiple
files with very different I/O patterns. These files from the same application
can usually be placed on different I/O subsystems. Therefore, our model focuses
on individual files instead of the whole application. Ideally, it predicts the
best I/O subsystem for each file and thus results in the best file placement
plan globally for the application. 

Since Lassen uses node-local BBs, the target is either using it or
not. So it should be a binary classification problem. We evaluate several different
classifiers and compare their test accuracies. We do 90-10 split on our dataset. We train the model with 90\% of the data, and then use the rest 10\%, which is unseen for the model, to get the test
accuracy by comparing predictions with the true labels. To mitigate the
variability from data split, we train and evaluate each classifier 10 times with
random data split every time. We report the test accuracy of the classifiers we
tried in the result section.

\subsection{Workflow for identifying best performing I/O sub-system}

\begin{figure}[h]
  \centering
    \includegraphics[width=\columnwidth]{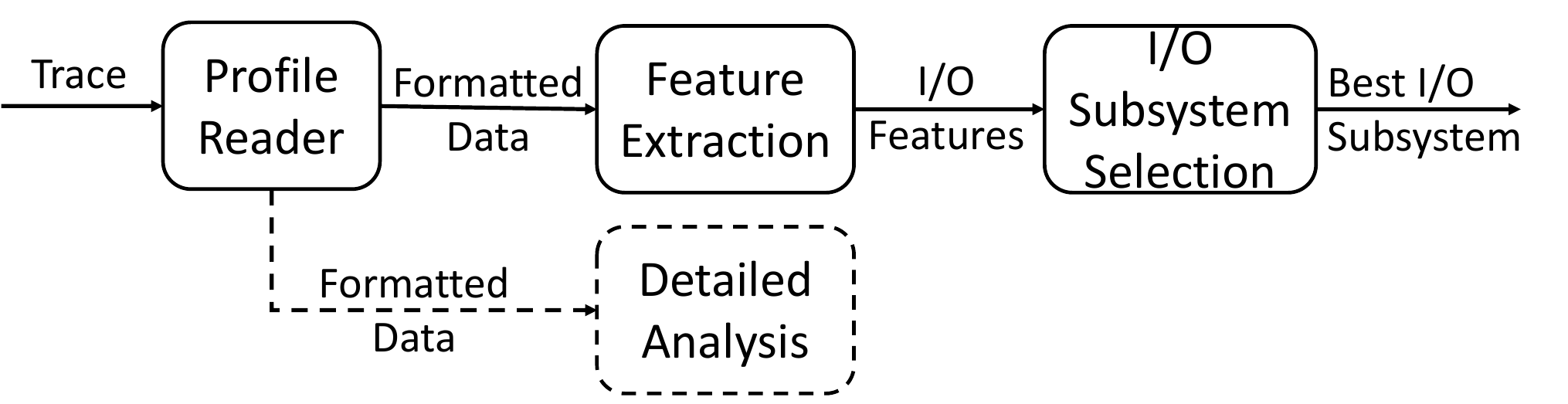}
  \caption{Workflow overview. Users provide the raw trace of their application
run and PrismIO will restructure it into a much more programmable format
(IOFrame). Then PrismIO extracts I/O features from IOFrame. These features are
provided as input to the model that decides the best I/O placement on I/O
subsystems to run this application.}
  \label{pipeline}
\end{figure}

We make the selection process efficient by developing a workflow. This
is because even though we obtained the approximation of the relationship with
the ML model, applying it to large-scale HPC applications with a large number
of different files is non-trivial. When users don't know the I/O features of an
application, they cannot apply the model. To solve this problem, we combine the
feature extraction functions of PrismIO with the model and make them into a
workflow. Figure \ref{pipeline} highlights the workflow design and demonstrates
how parts fit together in the workflow and their inputs/outputs. Given an
application, the user should only need to run it once on any machine and trace
it with Recorder.  Then the workflow would take the Recorder trace, extract I/O
features for each file, and feed them into the pre-trained model to predict the
best file placement, namely which file should be placed on which system for all
files. 

\begin{figure}[h]
  \centering
    \includegraphics[width=\columnwidth]{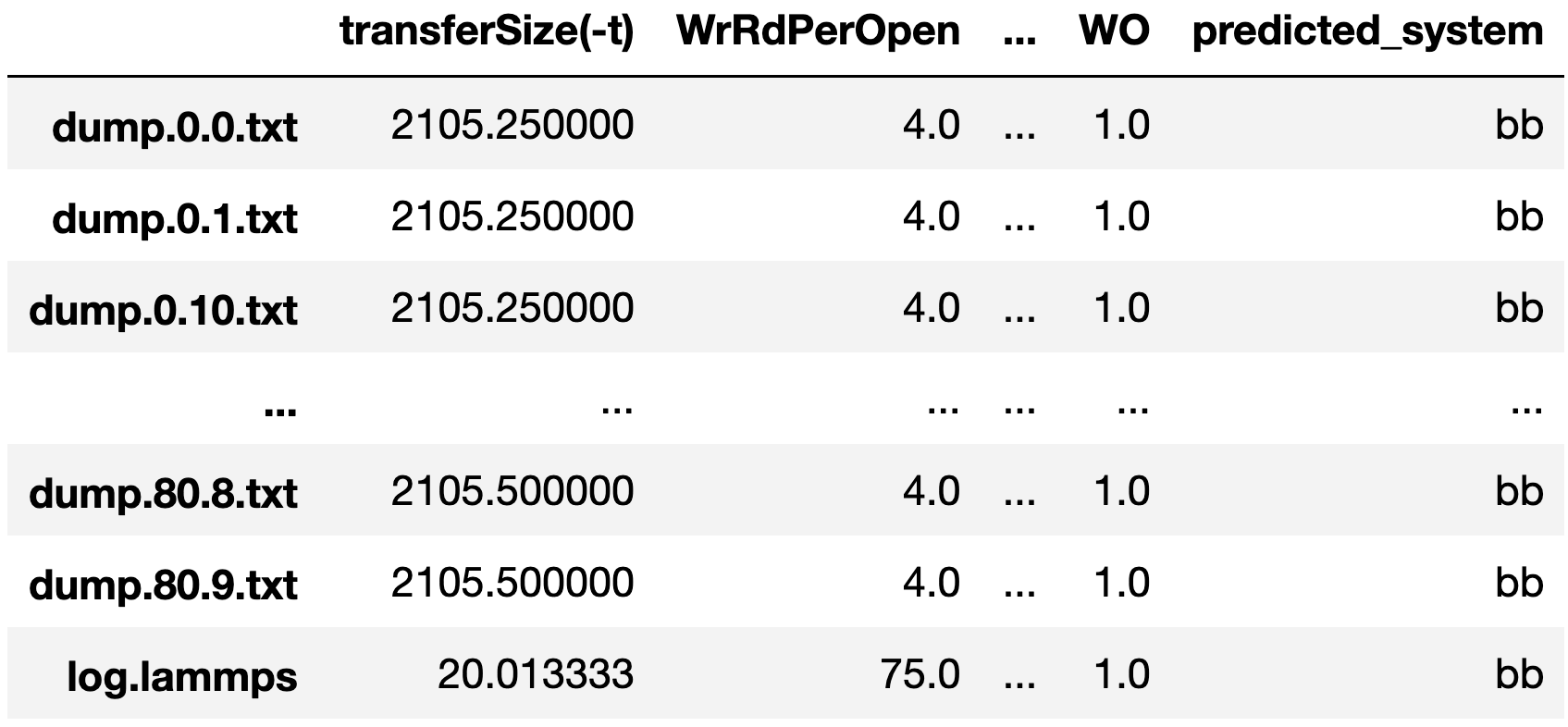}
\begin{lstlisting}
result = predict("/path/to/recorder/trace")
\end{lstlisting}
  \caption{A screenshot of a part of the DataFrame of prediction workflow
output of a sample trace. Each row corresponds to a file. We can easily
understand what are the I/O features of this file, and given those features,
where should we put the file. In this example, we know that all I/O to these
files are write, and there are 4 writes with an average size of 2105.25 bytes.
And based on this, all of them are predicted to perform better when placed on
BB.}
  \label{workflow-output}
\end{figure}

We have made the workflow a function called \texttt{predict} in PrismIO. Users
simply need to call it with the trace path. It groups the data by file names
and calls feature extraction APIs on each of them. It loads the model and
predicts whether to use BBs for each file. The I/O features and predictions are
organized as a dataframe indexed by file name. Figure \ref{workflow-output}
demonstrates a part of the workflow output of a sample trace. 

\section{Results}
In this section, we present a comprehensive evaluation of our ML models. First,
we report test accuracy for all classifiers we have tried on IOR benchmarking
data. Second, we evaluate our prediction workflow with four real applications.
We report the accuracy of the workflow on real applications.

\subsection{Comparing different ML models}

\begin{figure}[h]
    \centering
      \includegraphics[width=0.9\columnwidth]{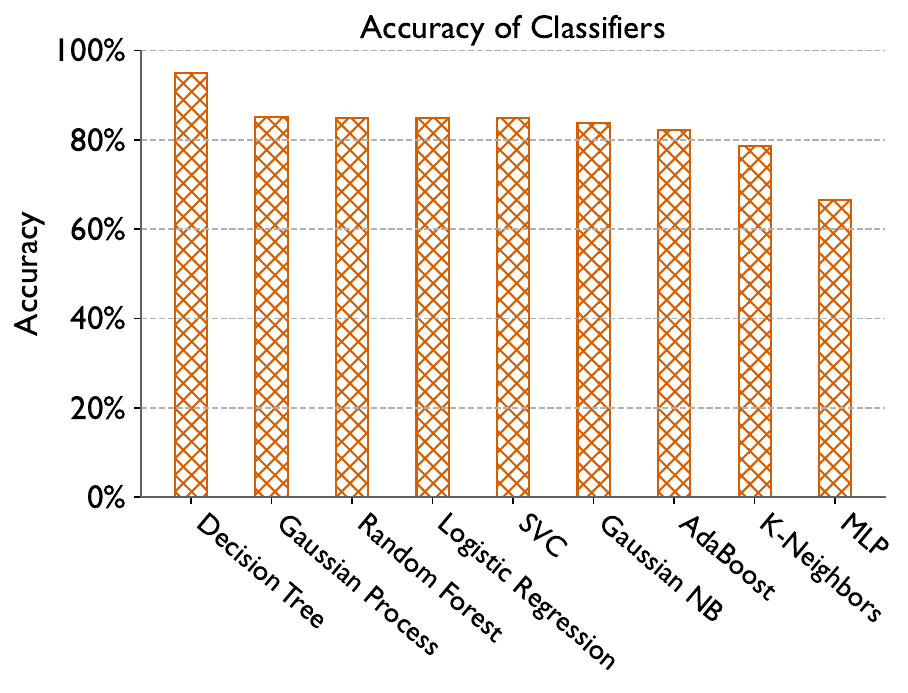}
    \caption{Accuracy of all tested classifiers. We model whether to put files
on BBs based on the I/O features of an application. We train classifiers with
IOR data produced in our experiment. The dataset contains 2967 samples and we
do a 90-10 split for training and testing. We do 10 random splits and evaluate
the model and take the average accuracy to minimize the possible impact from
certain data splits. We pick the classifier with the highest accuracy to use in
our model.}
    \label{classifier-accuracy}
\end{figure}

We use the percentage of correct prediction to evaluate our classifiers. In
other words, we compare the predictions with the ground truth and see what
percentage it predicts the correct answer. We use accuracy to refer to this.
For Lassen among the classifiers we tried, the Decision Tree classifier gives
the highest test accuracy, which is 94.47\% (Figure \ref{classifier-accuracy}).
Therefore, we decide to continue with the Decision Tree classifier.

\begin{figure}[h]
    \centering
      \includegraphics[width=0.9\columnwidth]{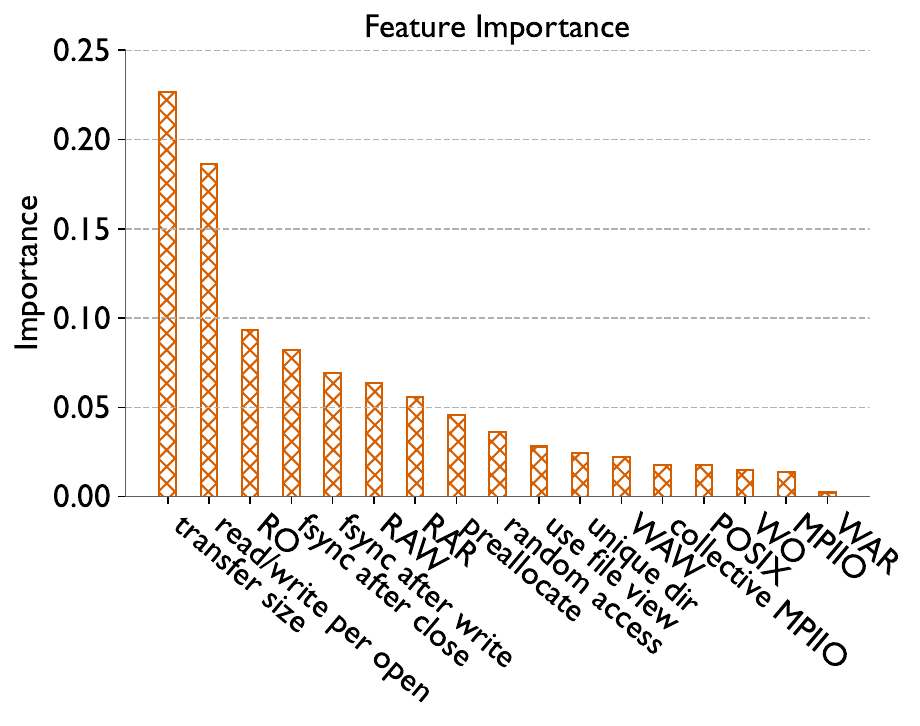}
    \caption{Importance of all features of Decision Tree classifier. The value
is between 0 and 1. A higher value means the feature is more important.  For
example, the feature importance of transfer size is 0.23. It is important in
deciding whether to use BB.}
    \label{feature-importance}
\end{figure}

The Decision Tree classifier has metrics for feature importance.  Figure
\ref{feature-importance} demonstrates the importance of all features. Since our
model predicts whether a combination of I/O features performs better on GPFS or
BB, higher feature importance for model prediction implies the feature can be a
key factor that distinguishes GPFS and BB. It may also imply underlying
bottlenecks of a certain system. On the other hand, features with very low
importance may be insignificant for prediction and accuracy. Therefore,
features are selected after model selection by eliminating the least important
features recursively until the accuracy significantly drops. After selecting
the classifier and feature set, the chosen classifier is trained using the same
data. This training is offline and the trained model can be loaded at runtime.

\subsection{Model evaluation using production applications}

In section V, we presented our test of the model on unseen IOR runs and got a
94.47\% accuracy for Lassen. However, IOR is not a real application and we
directly know the I/O features from the configurations of IOR runs. To further
evaluate how our prediction workflow works on real applications, we test it
with four unseen applications: LAMMPS, CM1, Paradis, and HACC-IO. LAMMPS is a
molecular dynamics simulator, CM1 is an extreme climate simulator, Paradis is a
dislocation dynamics simulator, and HACC-IO is the I/O proxy application of
HACC, which is a cosmology simulator. 

First, we run these applications on both Lassen GPFS and BBs and trace them
with Recorder. Although we only need the trace on any one of the systems to
make the prediction, we need both to get the ground truth. We run them with
configurations from example use cases provided in the documentation or
repository of these applications. Second, for each run, we apply our prediction
workflow on either the GPFS trace or BB trace. As discussed in section V, it
predicts which I/O subsystem to use for each file. Third, we get the ground
truth of whether to use BBs for each file by comparing its I/O bandwidth on
GPFS and BBs. We apply the PrismIO \texttt{io\_bandwidth} API discussed in
section III to get the I/O bandwidths of each file. Finally, we compare the
predicted system with the ground truth to get accuracy.

We get 121 samples in our test data set. Since our model predicts whether to
use BBs for a file based on its I/O features, each individual file produced by
runs should be a sample. However, although these applications write a large
number of files, many of them have exactly the same I/O features. For example,
one LAMMPS run in our experiment has 386 different files, but most of them are
dump files.  LAMMPS does multiple dumps and it writes files in the same way for
each dump.  Such files will have exactly the same I/O features and the
prediction for them will also be the same, meaning they cannot be used as
different samples.  Therefore, we create more application runs by varying their
input configurations such as problem size, the number of processes, interfaces,
etc. They will have more files with different I/O features and thus produce
more samples. We group files with the same I/O features and treat them as a
single sample.  We use average bandwidth to figure out the ground truth for the
group.

\begin{figure}[h]
  \centering
    \includegraphics[width=0.9\columnwidth]{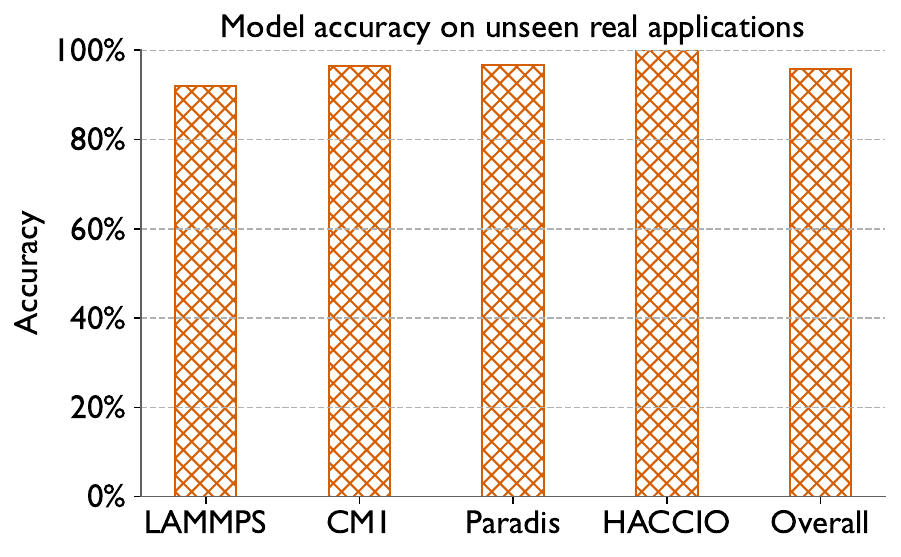}
  \caption{Accuracy of our prediction workflow on four real applications. We
run and trace these four applications. We feed their trace into our workflow
and get the predicted better system. For prediction, users only need one trace
on any system. But since we need to use ground truth to compute the accuracy,
we run them on both systems and compute the bandwidth of each file from their
trace. Overall we achieve a good accuracy of 95.86\%}
  \label{application-accuracy}
\end{figure}

Figure \ref{application-accuracy} demonstrates the test accuracy. We achieve
95.86\% accuracy on the whole test data set. For individual applications,
we experimented with seven different HACC-IO cases and achieved 100\% accuracy
because it has quite simple I/O patterns. We experimented with two Paradis
cases, three CM1 cases, and three LAMMPS cases to cover different I/O features.
We achieved 96.67\%, 96.55\%, and 92\% accuracy for Paradis, CM1, and LAMMPS,
respectively.

\section{Conclusion}
In this paper, we presented a prediction workflow that predicts whether to put
files on BBs or not based on the I/O characteristics of an application. We did
an experiment that covered a variety of I/O characteristics combinations using
IOR.  We trained and tested our model using the experiment data and it achieved
94.47\% accuracy on unseen IOR data. Moreover, we developed feature extraction
functions to apply our model to applications when we don't know their I/O
characteristics.  we demonstrated use cases of the prediction workflow of runs
of four real applications.  It achieved 95.86\% overall accuracy on those runs.

Besides the prediction workflow, we presented PrismIO, a Python-based library
that enables programmatic analysis. It provides APIs for users to do detailed
analyses efficiently. With the tool, we demonstrated two case studies where we
utilized PrismIO to conduct detailed analyses.  We explained the causes of two
I/O performance issues on Lassen GPFS and BBs and made some empirical
conclusions more complete.  The case studies demonstrate the potential of
PrismIO in making data analytics quick and convenient for the HPC user.

In the future, we plan to apply the same methodology to build prediction
workflow for other platforms that have BBs such as Summit. Moreover, the
current feature extraction APIs are not fast enough, therefore, we plan to
improve their performance by designing better algorithms and creating APIs that
extract multiple features at the same time. We believe that will make our
prediction workflow more efficient for large application runs.

\section*{Acknowledgment}
This work was performed under the auspices of the U.S.~Department of Energy
by Lawrence Livermore National Laboratory under Contract DE-AC52-07NA27344
(LLNL-CONF-858971). This material is based upon work supported by the
U.S.~Department of Energy, Office of Science, Office of Advanced Scientific
Computing Research under the DOE Early Career Research Program.

\bibliographystyle{IEEEtran}
\bibliography{./bib/cite,./bib/pssg}

\end{document}